\documentclass[aps,prl,floatfix,showpacs,twocolumn,superscriptaddress]{revtex4}

\usepackage{color}
\usepackage{bm}
\usepackage{amssymb}
\usepackage{amsmath}
\usepackage{amstext}
\usepackage{graphics}
\usepackage{epsfig}
\usepackage{placeins}
\usepackage{psfrag}
\usepackage{subfigure}
\usepackage{times}

\newcommand{\bv}[1]
{
\left(
\begin{matrix}
#1
\end{matrix}
\right)
}

\begin{document}

\title{Proposal of a robust measurement scheme for the non-adiabatic spin torque using the displacement of magnetic vortices}

\author{Benjamin Kr\"uger}

\affiliation{I. Institut f\"ur Theoretische Physik, Universit\"at Hamburg,
  Jungiusstr. 9, 20355 Hamburg, Germany}

\author{Massoud Najafi}

\affiliation{Arbeitsbereich Technische Informatik Systeme, Universit\"at Hamburg, Vogt-K\"olln-Str. 30, 22527 Hamburg, Germany}

\author{Stellan Bohlens}

\affiliation{I. Institut f\"ur Theoretische Physik, Universit\"at Hamburg,
  Jungiusstr. 9, 20355 Hamburg, Germany}

\author{Robert Fr\"omter}

\affiliation{Institut f\"ur Angewandte Physik, Universit\"at Hamburg,
  Jungiusstr. 11, 20355 Hamburg, Germany}

\author{Dietmar P. F. M\"oller}

\affiliation{Arbeitsbereich Technische Informatik Systeme, Universit\"at Hamburg, Vogt-K\"olln-Str. 30, 22527 Hamburg, Germany}

\author{Daniela Pfannkuche}

\affiliation{I. Institut f\"ur Theoretische Physik, Universit\"at Hamburg,
  Jungiusstr. 9, 20355 Hamburg, Germany}

\begin{abstract}
The strength of the non-adiabatic spin torque is currently under strong debate, as its value differs by orders of magnitude as well in theoretical predictions as in measurements. Here, a measurement scheme is presented that allows to determine the strength of the non-adiabatic spin torque accurately and directly. Analytical and numerical calculations show that the scheme allows to separate the displacement due to the Oersted field and is robust against uncertainties of the exact current direction.
\end{abstract}


\maketitle

A spin-polarized current flowing through a ferromagnetic sample interacts with the magnetization and exerts a torque on the local magnetic moments. For conduction electron spins that follow the local magnetization adiabatically it has been shown that the interaction via spin transfer can be described by adding a current-dependent term to the Landau-Lifshitz-Gilbert equation.\cite{PhysRevB.57.R3213} This equation has been extended by an additional term that takes the non-adiabatic influence of the itinerant spins into account.\cite{PhysRevLett.93.127204} Theoretically, several mechanisms have been proposed as the origin of the non-adiabatic spin torque, leading to different orders of magnitude for its strength.\cite{PhysRevLett.93.127204,kohno,tserkovnyak:144405,duine:214420,tatara} Thus a precise measurement of the non-adiabatic spin torque is necessary to give insight into its microscopic origin. A determination of its strength is further important for a reliable prediction of the current-driven domain-wall velocity.\cite{PhysRevLett.93.127204} Currently measured values of the non-adiabatic spin torque for permalloy differ by one order of magnitude,\cite{hayashi:197207,meier:187202,heyne:066603,ThomasNAT06} thus the strength of the non-adiabatic spin torque is under strong debate. In these experiments the observed motion of a domain wall was compared with micromagnetic simulations to determine the non-adiabatic spin torque. This analysis is highly susceptible to surface roughness and Oersted fields.

Due to its high symmetry and spacial confinement a vortex in a micro- or nanostructured magnetic thin-film element is a promising system for the investigation of the spin-torque effect.\cite{sp5,bolte:176601,kasai:237203} Vortices are formed when the in-plane magnetization curls around a center region. In this few-nanometer-large center region, called the vortex core, the magnetization turns out of plane to minimize the exchange energy. There are four different ground states of a vortex. These states are labeled by the direction of the out-of-plane magnetization, called polarization $p$, and the sense of rotation of the in-plane magnetization, called chirality $c$. Polarizations of $p = 1$ and $p = -1$ denote a core that points parallel or antiparallel to the $z$ axis, respectively. A chirality of $c = 1$ denotes a counterclockwise curling of the in-plane magnetization while $c = -1$ denotes a clockwise curling.

It is known that vortices are displaced from their equilibrium position when excited by spin-polarized electric current pulses.\cite{bolte:176601,kasai:237203,yamada,krueger:224426,kruger:07A501,guslienko:067205,lee:192513,lee:014405,shibata:020403} The spatial confinement of the vortex core within the film element yields an especially accessible system for measurements with scanning probe techniques, such as soft x-ray microscopy, x-ray photoemission electron microscopy, or scanning electron microscopy with polarization analysis. An analytical solution of the extended Landau-Lifshitz-Gilbert equation shows that for a current-driven vortex the forces due to the adiabatic and the non-adiabatic spin torque are perpendicular to each other.\cite{krueger:224426}

In this work we present a scheme which allows to measure the contributions due to the adiabatic spin torque, the non-adiabatic spin torque, and the Oersted field separately. It bases upon analytical calculations~\cite{krueger:224426} and overcomes the two main difficulties that occur in an experiment. The first problem arises from an additional vortex displacement due to the Oersted field accompanying the current flow.\cite{bolte:176601} This displacement is comparable in size to the displacement due to the non-adiabatic spin torque and both displacements point in the same direction.\cite{krueger:224426} Thus, the unknown contribution of the Oersted field has to be separated from the measured signal. The second problem is the exact determination of the displacement angle. Since the displacement due to the adiabatic spin torque is about one order of magnitude larger than the displacement due to the non-adiabatic spin torque, a small uncertainty in the direction of the current through the sample would cause large errors in the determination of the non-adiabatic spin torque parameter. To test the applicability of our analytical findings they are applied to vortex displacements obtained from three-dimensional micromagnetic simulations.

For the analytical calculations we start from a modified version of the Thiele equation \cite{PhysRevLett.30.230,0295-5075-69-6-990}

\begin{equation}
\vec F + G_0 \vec e_z \times ( \vec v_c + b_j \vec j) + D_{\Gamma} \alpha \vec v_c + D_0 \xi b_j \vec j = 0
\end{equation}
that takes deformation of the vortex into account.\cite{sup} Here $v_c$ is the velocity of the vortex core, $\alpha$ is the Gilbert damping, $\xi$ the strength of the non-adiabaticity, $\vec F$ the force on the vortex, $G_0$ the $z$ component of the gyrovector, and $D_0$ the diagonal element of the dissipation tensor. The coupling constant $b_j = P \mu_{\mbox{\scriptsize B}}/(e M_{\mbox{\scriptsize s}})$ between the current and the magnetization depends on the saturation magnetization $M_{\mbox{\scriptsize s}}$ and the spin polarization $P$. The assumption of a magnetization pattern which rigidly gyrates holds true only for the small vortex core. Due to the spatial confinement the remaining part of the vortex has to deform while the core is moving. $D_{\Gamma} < D_0$ is a phenomenological parameter that takes into account a reduced dissipation due to this deformation.\cite{sup}

We will investigate a square thin-film element with a homogeneous current flowing in $x$ direction. The accompanying Oersted field is accounted for by a homogeneous field $H$ in $y$ direction. For small displacements of the vortex core from its equilibrium position the demagnetization energy can be expanded up to second order in the core displacement $\vec R = (X,Y)$. The force on the vortex is then given by~\cite{krueger:224426}
\begin{equation}
\vec F = - \bv{\mu_0 M_{\mbox{\scriptsize s}} H l d c + m \omega_{\mbox{\scriptsize r}}^2 X\\ m \omega_{\mbox{\scriptsize r}}^2 Y},
\end{equation}
with the lateral extension $l$, and thickness $d$ of the system. The factor $m \omega_{\mbox{\scriptsize r}}^2$ parameterizes the confining potential.\cite{krueger:224426}

This equation can be solved for harmonic excitations of the form $\vec H(t) = H_0 e^{i \Omega t} \vec e_y$ and $\vec j(t) = j_0 e^{i \Omega t} \vec e_x$. After all transient states are damped out, the position of the vortex core is given by
\begin{widetext}
\begin{equation}
\bv{X\\ Y} = -\frac{e^{i \Omega t}}{\omega^2 + (i \Omega + \Gamma)^2} \bv{\tilde j & \tilde H c p + \left| \frac{D_0}{G_0} \right | p\xi \tilde j\\ -\tilde H c p - \left| \frac{D_0}{G_0} \right | p\xi \tilde j& \tilde j} \bv{ \frac{\omega^2}{\omega^2 + \Gamma^2} i \Omega
\\ \omega p + \frac{\omega \Gamma}{\omega^2 + \Gamma^2} i \Omega p},
\label{eq_solution}
\end{equation}
\end{widetext}
with $\tilde H = \gamma H_0 l /(2 \pi)$, the gyromagnetic ratio $\gamma$, $\tilde j = b_j j_0$, the free oscillation frequency of the vortex $\omega$, and its damping $\Gamma$.\cite{sup}

From Eq.~(\ref{eq_solution}) it is obvious that an Oersted field has the same influence on the vortex as the non-adiabatic spin torque. Thus the presence of an Oersted field can disturb the measurement of the non-adiabatic spin torque. In experiments the coordinate system is given by the sample axis. A small uncertainty of the direction of the current flow, e.g. due to a rotation or imperfections of the sample, yields a mixing of the displacement components, resulting from the adiabatic spin torque and the smaller non-adiabatic spin torque, relative to the sample axis. This mixing causes a large error in the measurement of the displacement originating from the non-adiabatic spin torque.

An excitation with a direct current causes a displacement of the vortex core to a new steady-state position.\cite{resonant} A benefit is that a direct current allows for a measurement with a non-time-resolving technique. The displacement of a vortex driven by a direct current is given by
\begin{equation}
\vec R_c^p(j) = -\frac{\omega}{\omega^2 + \Gamma^2} \bv{ \tilde H c + \left| \frac{D_0}{G_0} \right | \xi \tilde j\\ \tilde j p}.
\label{eq_solution_dc}
\end{equation}

\begin{figure}
\includegraphics[angle = 270, width = 1.0\columnwidth]{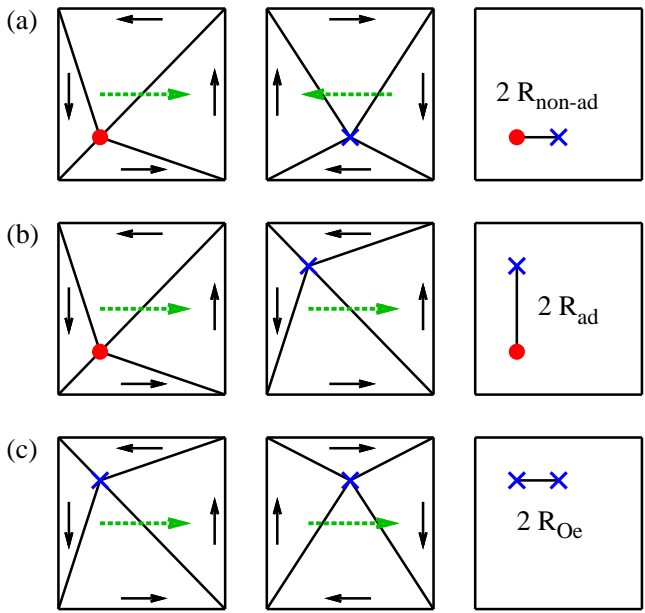}
\caption{Scheme of the determination of the three different contributions to the vortex displacement. By measuring the distance between the positions of two different vortices it is possible to separate the displacements (a) due to the non-adiabatic spin torque, (b) the adiabatic spin torque, and (c) the Oersted field. Points and crosses denote cores with positive and negative polarization, respectively. The in-plane magnetization is denoted by the solid arrows. The dashed arrows denote the current direction. For the sake of illustration the displacements are exaggerated.\label{diff}}
\end{figure}

\begin{figure}
\includegraphics[angle = 270, width = 1.0\columnwidth]{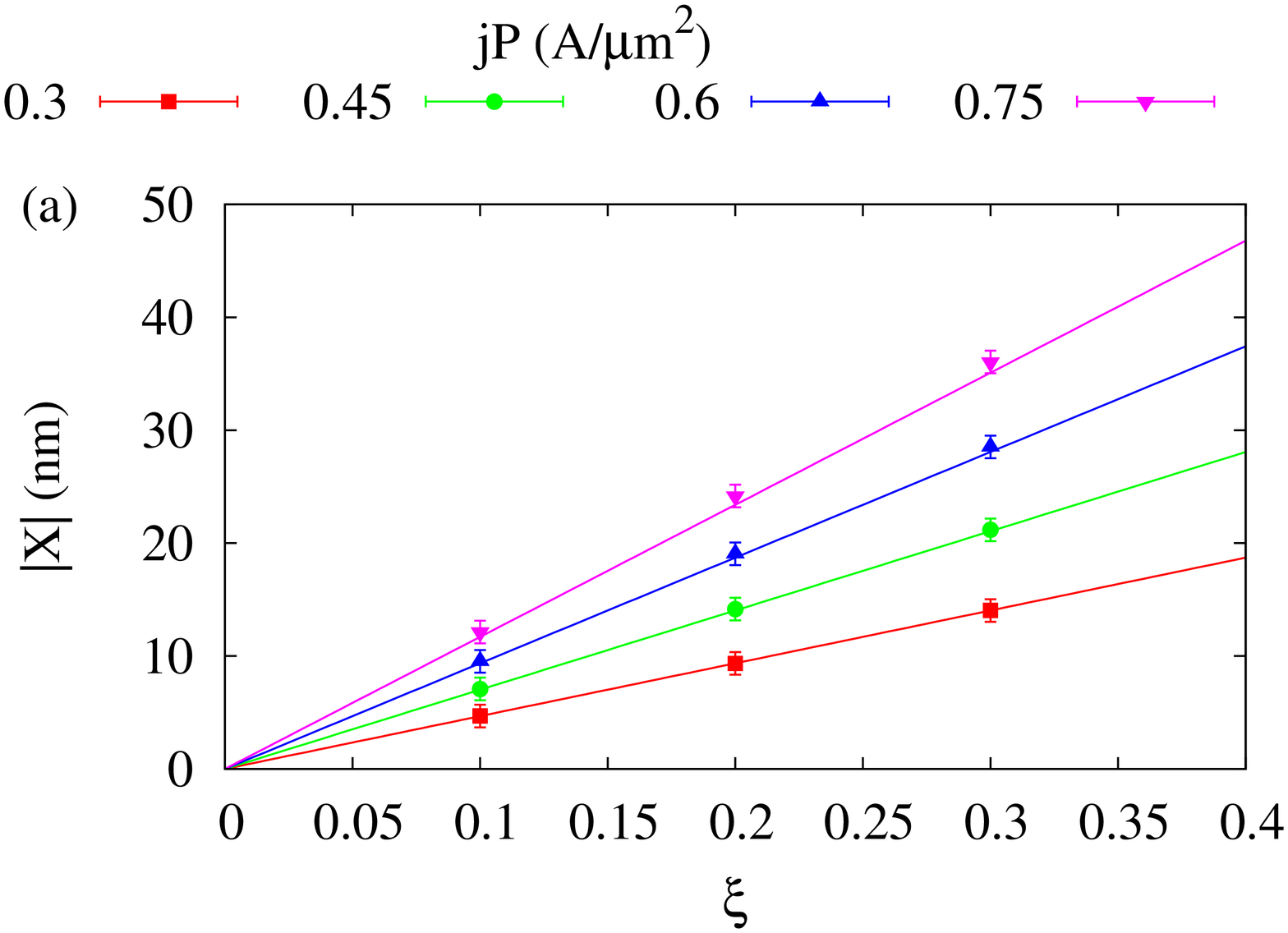}
\includegraphics[angle = 270, width = 1.0\columnwidth]{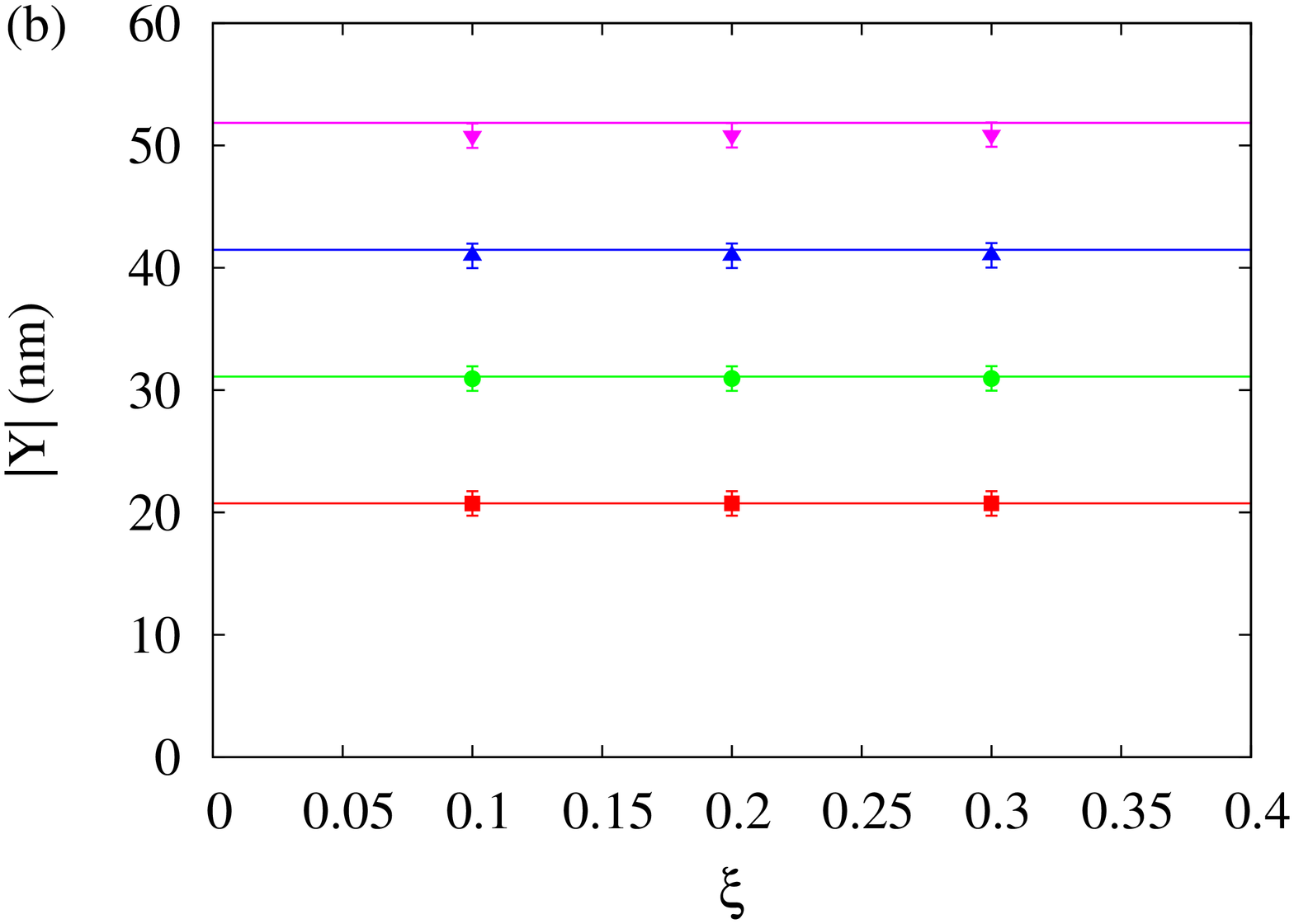}
\caption{Numerically calculated displacement of the vortex core due to a direct spin-polarized current of density $jP$ in the absence of an Oersted field. (a) The displacement parallel to the current is proportional to $\xi$. (b) The displacement perpendicular to the current is independent of $\xi$. The lines are fits with the linear model in Eq.~(\ref{eq_solution_dc}). For large current densities small non-linear effects can be seen.\label{displacement}}
\end{figure}

The sign of the displacement induced by the Oersted field depends on the chirality of the vortex, while the displacement due to the adiabatic spin torque is determined by the polarization.\cite{shibata:020403} The non-adiabatic spin torque causes a displacement that is independent of the vortex properties $p$ and $c$. Vortices with different $p$ and $c$ values can be achieved by remagnetizing the sample. Comparing the displacement of three vortices with different polarizations and chiralities it is therefore possible to separate the contributions of all three forces to the displacement of the vortex. From Eq. (\ref{eq_solution_dc}) we find
\begin{subequations}
\begin{align}
2 R_{\mbox{\scriptsize non-ad}} & = 2 \left| \frac{\omega \xi \tilde j}{\omega^2 + \Gamma^2} \frac{D_0}{G_0} \right | = \left| \vec R_{c}^{p}(j) - \vec R_{-c}^{-p}(-j) \right|
\label{eq_non-ad}\\
2 R_{\mbox{\scriptsize ad}} & = 2 \left| \frac{\omega \tilde j}{\omega^2 + \Gamma^2} \right | = \left| \vec R_{c}^{p}(j) - \vec R_{c}^{-p}(j) \right|
\label{eq_ad}\\
2 R_{\mbox{\scriptsize Oe}} & = 2 \left| \frac{\omega \tilde H}{\omega^2 + \Gamma^2} \right | = \left| \vec R_{c}^{-p}(j) - \vec R_{-c}^{-p}(j) \right|.
\label{eq_Oe}
\end{align}
\end{subequations}
These equations are schematically illustrated in Fig.~\ref{diff}. From Eqs.~(\ref{eq_non-ad}) and (\ref{eq_ad}) it is possible to determine the non-adiabaticity parameter as
\begin{equation}
\xi = \frac{2 R_{\mbox{\scriptsize non-ad}}}{2 R_{\mbox{\scriptsize ad}}} \left| \frac{G_0}{D_0} \right | = \frac{\left| \vec R_{c}^{p}(j) - \vec R_{-c}^{-p}(-j) \right|}{\left| \vec R_{c}^{p}(j) - \vec R_{c}^{-p}(j) \right|} \left| \frac{G_0}{D_0} \right |.
\label{eq_xi}
\end{equation}
Since this equation is independent of the strength of the Oersted field, the angle of the sample, the frequency $\omega$, the damping $\Gamma$, and the parameter $D_{\Gamma}$, it yields the sought measurement scheme. With this scheme a direct determination of $\xi$ is accessible. Only one micromagnetic simulation for the determination of $|D_0/G_0|$ is necessary since $|D_0/G_0|$ is independent of $\xi$ and $j$.

\begin{figure}
\includegraphics[angle = 270, width = 1.0\columnwidth]{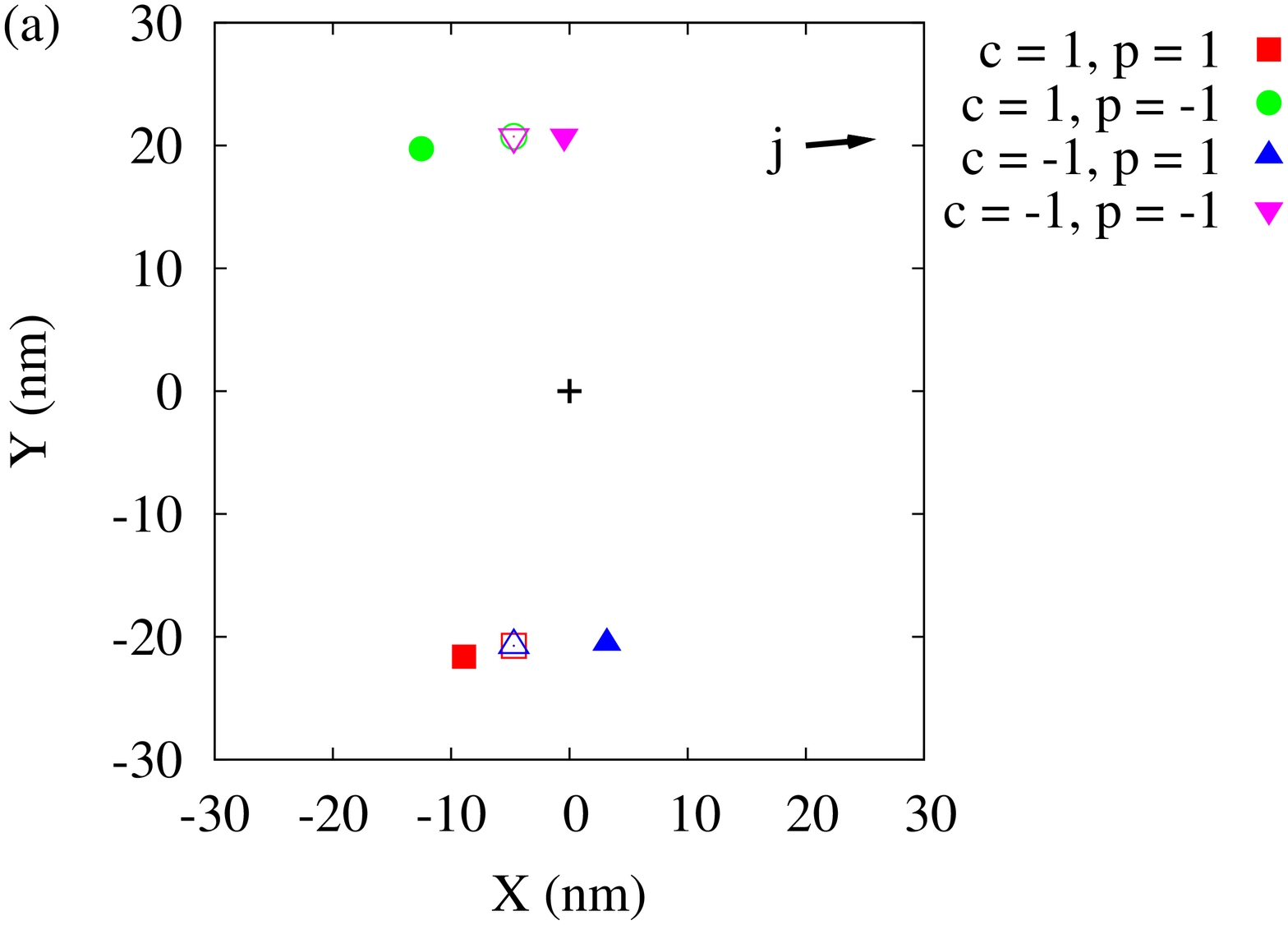}
\includegraphics[angle = 270, width = 1.0\columnwidth]{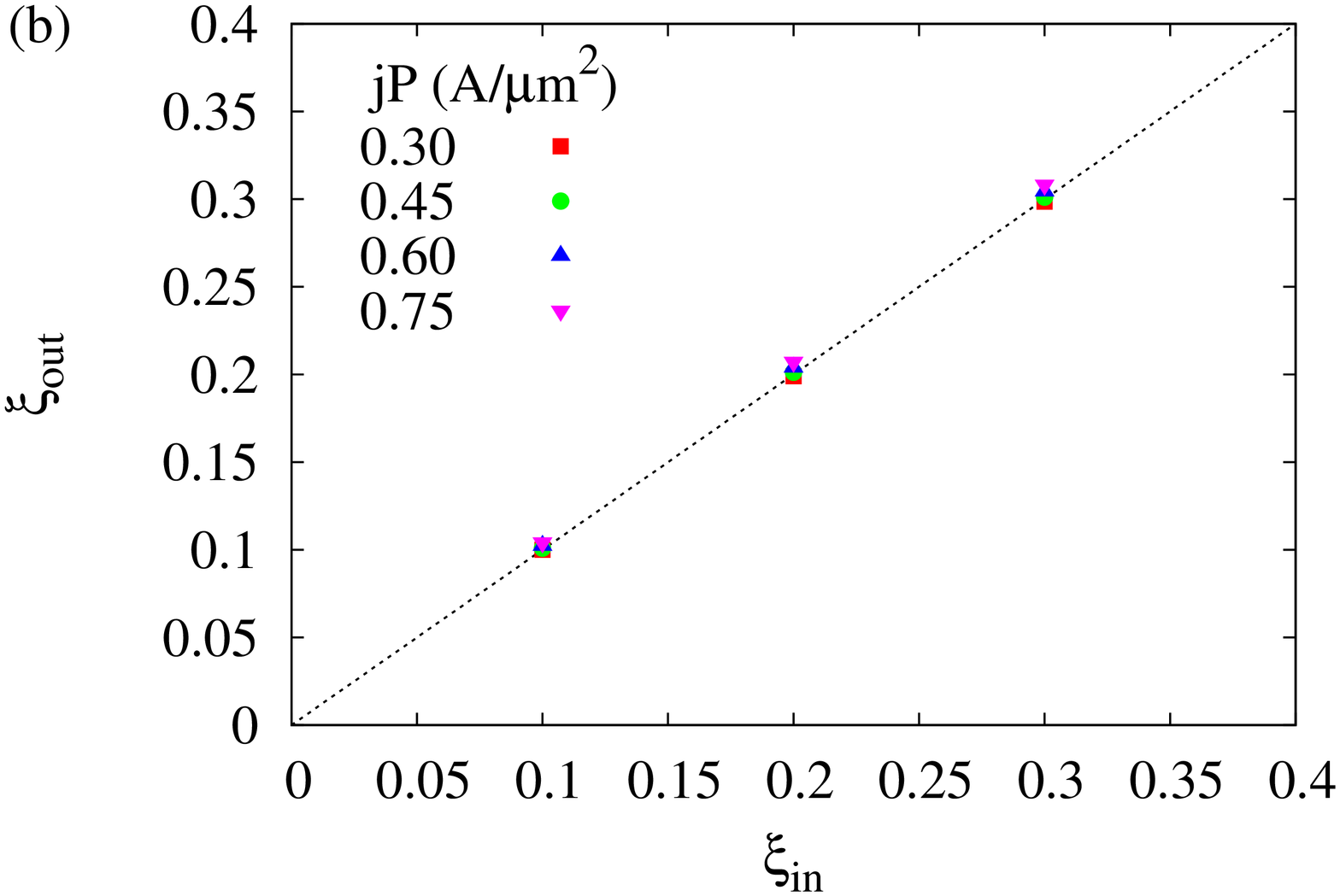}
\caption{(a) Position of the vortex core displaced by a spin-polarized direct current of density $jP = 3 \cdot 10^{11}$~A/m with a non-adiabatic spin-torque parameter of $\xi = 0.1$. The overlapping open symbols denote the positions for a current in exact $x$ direction without Oersted field. The closed symbols denote the positions with an applied Oersted field and a rotation of the sample by 5 degrees around its midpoint (plus). For the latter case the direction of the current is denoted by the arrow. (b) Results for $\xi_{\mbox{\scriptsize out}}$ derived from the positions shown in (a) using Eq.~(\ref{eq_xi}) for different current densities. $\xi_{\mbox{\scriptsize in}}$ is the value of the non-adiabaticity parameter that was used for the simulations.\label{positions}}
\end{figure}

Micromagnetic simulations of the experimental setup allow to determine the positions of the vortex core with a precise knowledge of the micromagnetic parameters of the system. The simulations therefore allow to test the analytical results in Eqs.~(\ref{eq_solution_dc}) and (\ref{eq_xi}). For the simulations the material parameters of permalloy, i.e., a saturation magnetization of $M_{\mbox{\scriptsize s}} = 8 \cdot 10^5$~A/m and an exchange constant of $A = 1.3 \cdot 10^{-11}$~J/m, are used. Since we are interested only in the steady final position of the vortex we used a Gilbert damping of $\alpha = 0.5$ to ensure a fast damping of the transient states to reduce computation time. As a sample system we considered a square thin-film element of length $l = 500$~nm and thickness $d = 10$~nm with a cell size of 2~nm in the lateral directions and 10~nm perpendicular to the film. This system allows for a reasonable computation time. For the simulations we used our extended version of the Object Oriented Micromagnetic Framework.\cite{OOMMF,krueger:054421}

Figure \ref{displacement} shows the displacement of the vortex core in simulations without Oersted field. As predicted by Eq.~(\ref{eq_solution_dc}) the displacement in the direction of the current flow is proportional to $\xi$ and the displacement perpendicular to the current flow is independent of $\xi$. From these simulations the value $|D_0/G_0| = 2.26$ can be determined.

In experimental samples we are faced with an Oersted field and an uncertainty of the direction of the current flow. To mimic the Oersted field in the simulations we  applied an in-plane field perpendicular to the current. The strength of the field is given by $H/(jP) = 1$~nm. The uncertainty of the direction of the current flow was taken into account by rotating the sample by 5 degrees. Figure~\ref{positions}(a) shows the positions of the vortex core for both simulations. It becomes visible that the Oersted field and the rotation of the sample strongly shift the core positions, complicating the determination of $\xi$.

To test the analytical model we compared the non-adiabatic spin-torque parameter $\xi_{\mbox{\scriptsize in}}$ that was inserted into the simulations with the value $\xi_{\mbox{\scriptsize out}}$ that was calculated from Eq.~(\ref{eq_xi}) using the core positions. Here it is worth noting that the value of the Oersted field and the angle of the sample are not needed for the calculation of $\xi_{\mbox{\scriptsize out}}$. The results are shown in Fig.~\ref{positions}(b). It can be seen that all the perturbations that are inserted in the simulations can be effectively excluded by the analytical calculations.

In experimental samples we are also faced with the anisotropic magnetoresistance (AMR) effect that leads to inhomogeneous current paths, i.e., a higher current density in the vortex core. Simulations including these inhomogeneous current paths yield a small shift to lower values of $\xi_{\mbox{\scriptsize out}}$. This shift is up to 2~\% for an AMR ratio of 10~\%.

In the remaining part we will discuss the experimental accuracy in the determination of $\xi$ that can be achieved with the presented scheme. In experiments direct currents of densities up to $1.5 \cdot 10^{12}$~A/m$^2$ have been realized in permalloy on a diamond substrate.\cite{hankemeier:242503} Assuming a spin polarization of 0.5 we get a spin-polarized current density of $0.75 \cdot 10^{12}$~A/m$^2$, i.e., the maximum shown in Fig.~\ref{displacement}. This yields values of up to $\tilde j = 55$~m/s.

The displacements of the vortex in the numerically investigated samples are small compared to the experimental resolutions available. A larger displacement of the vortex can be achieved by increasing the lateral size of the structure. For example simulations of a square thin-film element of length $l = 5000$~nm and thickness $d = 10$~nm yielded values of $|D_0/G_0| = 3.82$ and  $\omega/(\omega^2 + \Gamma^2) = 1 \cdot 10^{-8}$~s. With these values Eq.~(\ref{eq_ad}) yields $2 R_{\mbox{\scriptsize ad}} = $1100~nm. We assume that the core position can be measured with a resolution of $\delta (2 R_{\mbox{\scriptsize non-ad}}) = 20$~nm. Equation~(\ref{eq_xi}) then yields that $\delta \xi = 0.005$ can be realized. This resolution ranges from 5~\% to 50~\% depending on the value of $\xi$.\cite{hayashi:197207,meier:187202,heyne:066603,ThomasNAT06} The resolution can be further increased by using thin-film elements with still larger lateral sizes.

In conclusion we present a robust and direct measurement scheme for the non-adiabatic spin torque using the displacement of magnetic vortices. This scheme allows to distinguish between the displacements of the vortex core due to the non-adiabatic spin torque, the adiabatic spin torque, and the Oersted field, independently of the direction of the current flow. We also showed that an inhomogeneous current due to the AMR effect can be neglected. The scheme thus allows a precise measurement of the non-adiabatic spin-torque parameter $\xi$.

\begin{acknowledgments}
Financial support by the Deutsche Forschungsgemeinschaft via SFB 668 "Magnetismus vom Einzelatom zur Nanostruktur" and via Graduiertenkolleg 1286 "Functional metal-semiconductor hybrid systems" is gratefully acknowledged.
\end{acknowledgments}


\begin{thebibliography}{29}
\bibitem[{res()}]{resonant}
\bibinfo{note}{A measurement of the non-adiabatic spin torque with a resonant
  excitation using an alternating current is not suitable, as small deviations
  of the exciting frequency from the resonance frequency cause strong
  deviations in the trajectory of the vortex.}
\end{thebibliography}

\clearpage
\section{Supporting Material}

For an analytical investigation the motion of the vortex is commonly described employing the Thiele equation.\cite{PhysRevLett.30.230,guslienko:8037,guslienko:067205,guslienko:022510,krueger:224426,lee:192513,lee:014405,kasai:237203} This equation is exact for the steady state motion of a non-deformable magnetization pattern. However, this assumption holds true only for the small vortex core. Due to the spacial restriction the vortex has to deform while the core is moving. This yields a small modification of the Thiele equation that is especially important for current-driven vortex motion in view of the non-adiabatic spin torque.

Here we present a modified Thiele equation that takes a deformation of the outer part of the vortex into account.

With the magnetization $\vec M$ and the magnetic field $\vec H$ a general version of the Thiele equation reads\cite{PhysRevLett.30.230}
\begin{equation}
\begin{split}
0 = & - \mu_0 \int dV \, \left[ (\vec \nabla \theta) \frac{\partial}{\partial \theta}+ (\vec \nabla \phi) \frac{\partial}{\partial \phi} \right] (\vec H \cdot \vec M)\\
& - \frac{M_s \mu_0}{\gamma} \int dV \, \sin(\theta) (\vec \nabla \theta \times \vec \nabla \phi) \times ( \vec v + b_j \vec j)\\
& - \frac{M_s \mu_0}{\gamma} \int dV \, (\vec \nabla \theta \vec \nabla \theta + \sin^2(\theta) \vec \nabla \phi \vec \nabla \phi) ( \alpha \vec v + \xi b_j \vec j)
\end{split}
\label{eq_Thile_start}
\end{equation}
with the saturation magnetization $M_S$, the gyromagnetic ratio $\gamma$, the current density $\vec j$, the Gilbert damping $\alpha$, the non-adiabaticity parameter $\xi$, and the coupling constant $b_j$ between current and magnetization. $\theta$ and $\phi$ are the out-of-plane and in-plane angle of the magnetization, respectively. The velocity $\vec v = \vec v (r)$ of the magnetization pattern may depend on the position. Assuming that the magnetization pattern does not deform the velocity is independent of the position. Then Eq. (\ref{eq_Thile_start}) can be written in its well known form\cite{0295-5075-69-6-990}
\begin{equation}
\vec F + \vec G \times ( \vec v_c + b_j \vec j) + D ( \alpha \vec v_c + \xi b_j \vec j) = 0
\label{eq_Thile}
\end{equation}
with the velocity $\vec v_c$ of the vortex core. Here
\begin{equation}
\vec F = - \mu_0 \int dV \, \left[ (\vec \nabla \theta) \frac{\partial}{\partial \theta}+ (\vec \nabla \phi) \frac{\partial}{\partial \phi} \right] (\vec H \cdot \vec M)
\end{equation}
denotes the force on the magnetization pattern.
\begin{equation}
\vec G = - \frac{M_s \mu_0}{\gamma} \int dV \, \sin(\theta) (\vec \nabla \theta \times \vec \nabla \phi) = G_0 \vec e_z
\end{equation}
is the gyrovector and
\begin{equation}
D = - \frac{M_s \mu_0}{\gamma} \int dV \, (\vec \nabla \theta \vec \nabla \theta + \sin^2(\theta) \vec \nabla \phi \vec \nabla \phi)
\end{equation}
is the diagonal dissipation tensor with $D_{xx} = D_{yy} = D_0$ and $D_{zz} = 0$. The term $D \alpha \vec v_c$ describes the dissipation of energy due to the changing magnetization.

\begin{figure}
\includegraphics[angle = 270, width = 0.6\columnwidth]{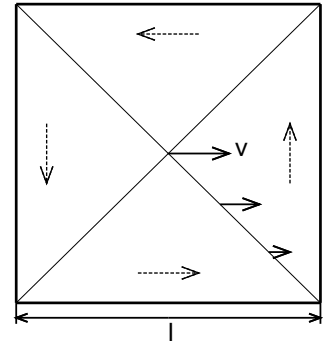}
\caption{Scheme of the magnetization (dashed arrows) in a square magnetic thin-film element with a vortex. The solid arrows denotes the velocity $v$ of the vortex core and of different points within the domain wall.\label{model}}
\end{figure}

\begin{figure}
\includegraphics[angle = 270, width = 1.0\columnwidth]{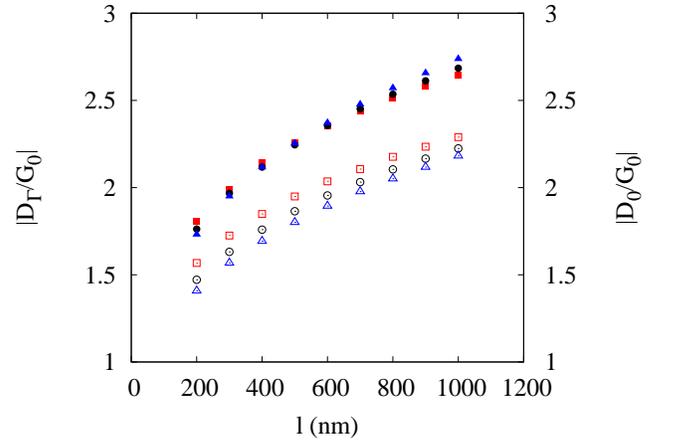}
\caption{Values of the strength $D_{\Gamma}$ of the dissipation (open symbols) and the strength $D_0$ of the non-adiabatic spin torque (closed symbols). The data for films of 10 nm, 20 nm, and 30 nm is denoted by squares, circles, and triangles, respectively.\label{D_Gamma}}
\end{figure}

\begin{figure}
\includegraphics[angle = 270, width = 1.0\columnwidth]{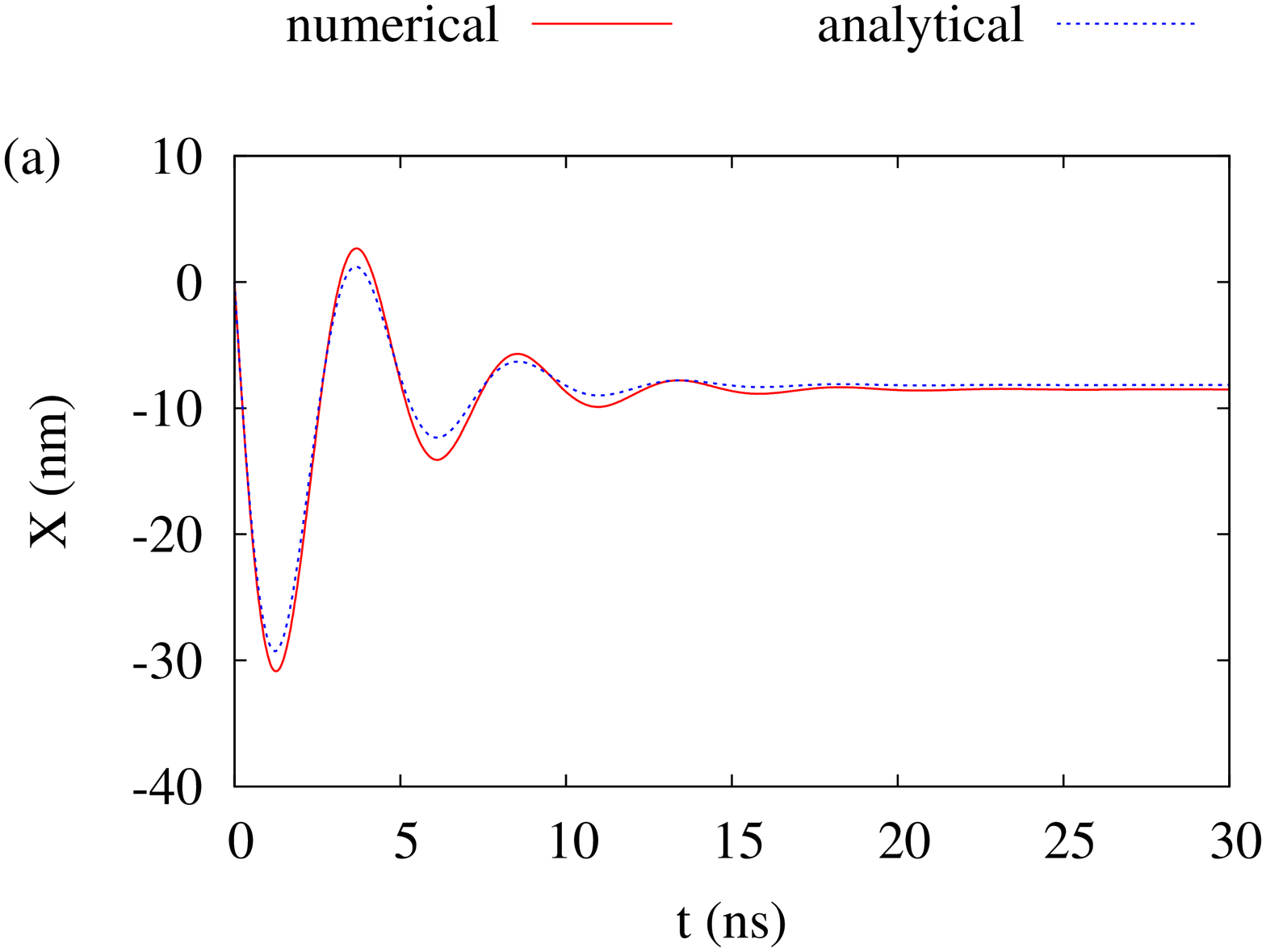}
\includegraphics[angle = 270, width = 1.0\columnwidth]{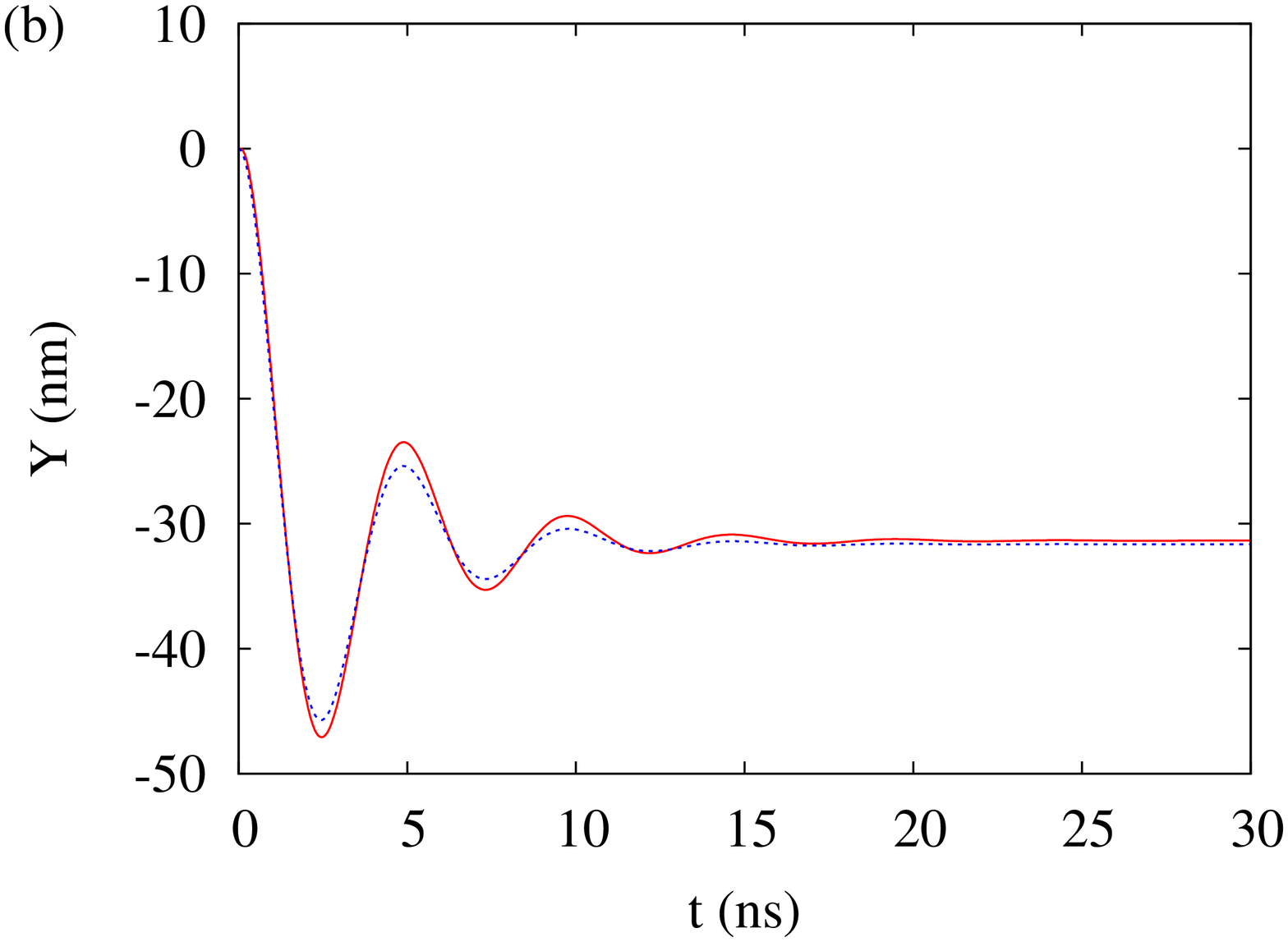}
\caption{Calculated position of a vortex core excited with a spin-polarized direct current of density $jP = 6 \cdot 10^{11}$~A/m$^2$ in a 1000~nm x 1000~nm square thin-film element. Shown is the $x$ position (a) and the $y$ position (b) versus time. A film thickness of 30~nm and $\xi = \alpha = 0.1$ was used. The solid (red) line is the vortex core position extracted from simulations. The dashed (blue) line is a fit with the theory based on the original Thiele equation.\label{fit1}}
\end{figure}

\begin{figure}
\includegraphics[angle = 270, width = 1.0\columnwidth]{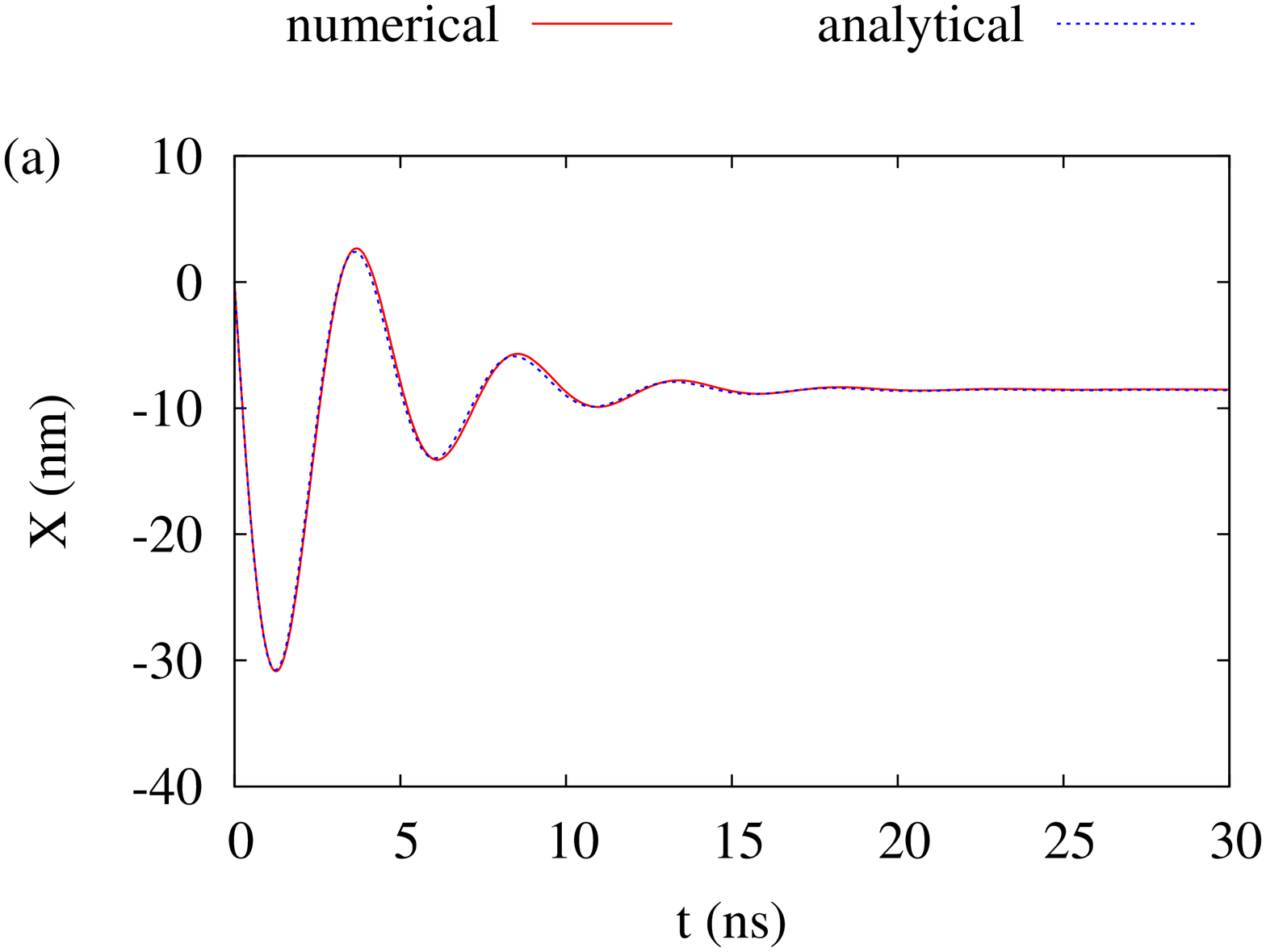}
\includegraphics[angle = 270, width = 1.0\columnwidth]{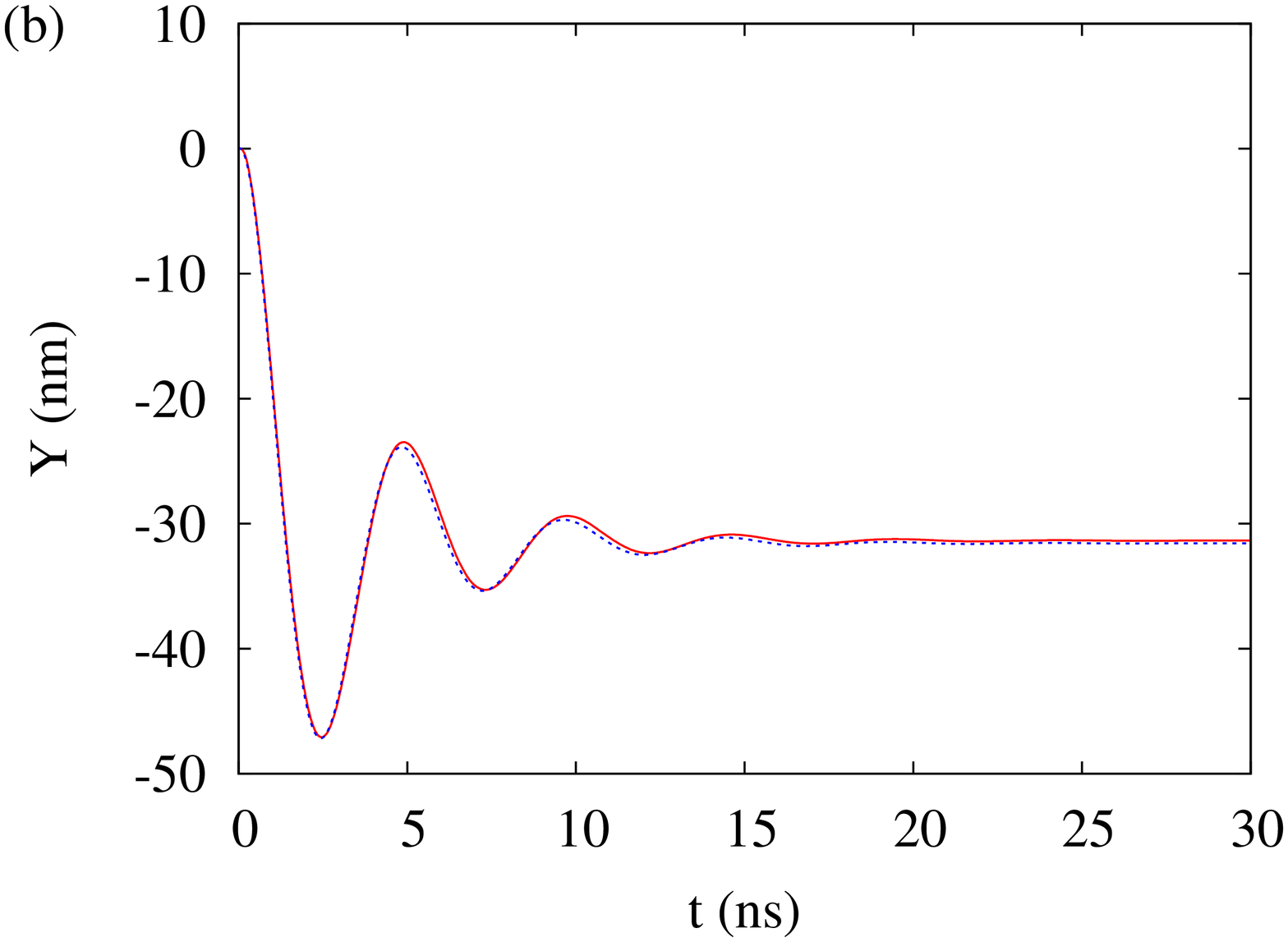}
\caption{Calculated position of a vortex core. The solid (red) line is the same as shown in Fig.~\ref{fit1}. The dashed (blue) line is a fit with the theory based on the modified Thiele equation.\label{fit2}}
\end{figure}

The integrand in the gyrovector is non-zero only in the small vortex core where the out-of-plane angle $\theta$ varies while the integrand in the dissipation tensor is also non-zero outside the core. Close to the boundaries of the sample the magnetization pattern moves slower compared to the center as it can be seen in Fig.~\ref{model}. Thus the velocity in the third term of Eq.~(\ref{eq_Thile_start}) depends on space. Aiming a similar form as in Eq.~(\ref{eq_Thile}) we replace the spatially dependent velocity $\vec v$ in the third term of Eq.~(\ref{eq_Thile_start}) by an effective value $\vec v_e$ which is independent of the position. This effective velocity occurs only in the third term as the second term is located at the vortex core. For a homogeneous current flow $b_j \vec j$ is constant over the sample. Thus we do not replace the current by an effective value. The equation then reads
\begin{equation}
\vec F + \vec G \times ( \vec v_c + b_j \vec j) + D_0 \alpha \vec v_e + D_0 \xi b_j \vec j = 0.
\label{eq_Thile2}
\end{equation}
The effective velocity $\vec v_e$ depends on the core position $\vec R = (X,Y)$ and the core velocity $\vec v_c$. For small deflections of the vortex core, i.e., small deformations of the vortex, $\vec v_e$ can be expanded in $\vec R$ and $\vec v_c$. For $\vec v_c = 0$ the magnetization is static and $\vec v_e = 0$. Thus the first non-vanishing term in the expansion is proportional to $\vec v_c$. Here and hereafter we write
\begin{equation}
\vec v_e = \frac{D_{\Gamma}}{D_0} \vec v_c.
\label{eq_diss_gamma}
\end{equation}
Since the effective velocity $\vec v_e$ is always smaller than the velocity $\vec v_c$ of the vortex core, the new constant $D_{\Gamma}$ is smaller than $D_0$. Inserting Eq.~(\ref{eq_diss_gamma}) in Eq.~(\ref{eq_Thile2}) yields a modified Thiele equation
\begin{equation}
\vec F + \vec G \times ( \vec v_c + b_j \vec j) + D_{\Gamma} \alpha \vec v_c + D_0 \xi b_j \vec j = 0.
\end{equation}

Employing the same conversions as used for the original Thiele equation \cite{krueger:224426} we find an expression for the velocity of the vortex core
\begin{equation}
\begin{split}
(G_0^2 + D_{\Gamma}^2 \alpha^2) \vec v_c & = \vec G \times \vec F - D_{\Gamma} \alpha \vec F - (G_0^2 + D_{\Gamma} D_0 \alpha \xi) b_j \vec j\\
 & + b_j \xi D_0 \vec G \times \vec j - b_j \alpha D_{\Gamma} \vec G \times \vec j.
\end{split}
\label{eq_velocity}
\end{equation}
We investigate a square thin-film element with a current in $x$ and a field in $y$ direction. With the stray-field energy for small deflections\cite{krueger:224426}
\begin{equation}
E_s = \frac{1}{2} m \omega_r^2 (X^2 + Y^2)
\end{equation}
and the total Zeeman energy\cite{krueger:224426}
\begin{equation}
E_z = \mu_0 M_s H l d c X
\label{eq_Zeeman_energy}
\end{equation}
we get a force of
\begin{equation}
\vec F = - \bv{\mu_0 M_s H l d c + m \omega_r^2 X\\ m \omega_r^2 Y}.
\end{equation}
Here $l$ and $d$ are the lateral extension and the thickness of the square, respectively. As for the original Thiele equation in the absence of current and field the excited vortex performs an exponentially damped spiral rotation around its equilibrium position. The free frequency
\begin{equation}
\omega = - \frac{p G_0 m \omega_r^2}{G_0^2 + D_{\Gamma}^2 \alpha^2}
\label{eq_omega_free}
\end{equation}
and the damping constant
\begin{equation}
\Gamma = - \frac{D_{\Gamma} \alpha m \omega_r^2}{G_0^2 + D_{\Gamma}^2 \alpha^2}
\label{eq_gamma}
\end{equation}
are slightly changed compared to their values derived from the homogeneous Thiele equation. In the following we express
\begin{equation}
D_{\Gamma} = \frac{\Gamma p G_0}{\omega \alpha}
\label{eq_omega_free_gamma}
\end{equation}
by the frequency and the damping constant. The velocity of the vortex then reads
\begin{widetext}
\begin{equation}
\bv{\dot X\\ \dot Y} = \bv{- \Gamma & -p \omega\\ p \omega & -\Gamma} \bv{X\\ Y} + \bv{\frac{p \omega \Gamma}{\omega^2 + \Gamma^2} \frac{\mu_0 M_s H l d c}{G_0} - \frac{\omega^2}{\omega^2 + \Gamma^2} b_j j - \frac{\Gamma \omega}{\omega^2 + \Gamma^2} \left| \frac{D_0}{G_0} \right | \xi b_j j\\ - \frac{\omega^2}{\omega^2 + \Gamma^2} \frac{\mu_0 M_s H l d c}{G_0} - \frac{p \omega \Gamma}{\omega^2 + \Gamma^2} b_j j + \frac{p \omega^2}{\omega^2 + \Gamma^2} \left| \frac{D_0}{G_0} \right | \xi b_j j}.
\label{eq_diff_velocity}
\end{equation}
This equation can be solved for harmonic excitations of the form $\vec H(t) = H_0 e^{i \Omega t} \vec e_y$ and $\vec j(t) = j_0 e^{i \Omega t} \vec e_x$. The solution for the vortex motion is then given by\cite{krueger:224426}
\begin{equation}
\bv{X\\ Y} = A \bv{i \\ p} e^{- \Gamma t + i \omega t} + B \bv{-i \\ p} e^{- \Gamma t - i \omega t} -\frac{e^{i \Omega t}}{\omega^2 + (i \Omega + \Gamma)^2} \bv{\tilde j & \tilde H c p + \left| \frac{D_0}{G_0} \right | p\xi \tilde j\\ -\tilde H c p - \left| \frac{D_0}{G_0} \right | p\xi \tilde j& \tilde j} \bv{ \frac{\omega^2}{\omega^2 + \Gamma^2} i \Omega
\\ \omega p + \frac{\omega \Gamma}{\omega^2 + \Gamma^2} i \Omega p},
\label{eq_solution_sup}
\end{equation}
\end{widetext}
with $\tilde H = \gamma H_0 l/(2 \pi)$ and $\tilde j = b_j j_0$. The first two terms with prefactors A and B are exponentially damped and depend on the starting configuration.

The values of $D_{\Gamma}$ and $D_0$ can be determined by micromagnetic simulations. For these simulations we used our extended version of the Object Oriented Micromagnetic Framework (OOMMF) that includes the adiabatic and non-adiabatic spin torque.\cite{PhysRevLett.93.127204,OOMMF,krueger:054421} The position of the vortex core was defined as the point with the maximum out-of-plane magnetization. To determine this maximum, the simulation cell with maximum out-of-plane magnetization and its next neighbors are interpolated with a polynomial of second order. For the simulations the material parameters of permalloy, i.e., a saturation magnetization of $M_s = 8 \cdot 10^5$~A/m and an exchange constant of $A = 1.3 \cdot 10^{-11}$~J/m where used.

For the determination of $D_{\Gamma}$ the vortex was excited by a magnetic field pulse. The subsequent oscillation was then fitted with the first two terms in Eq. (\ref{eq_solution_sup}). $D_{\Gamma}$ can then be determined from Eq. (\ref{eq_omega_free_gamma}). Finally the value of $D_0$ was determined by fitting an excitation with a direct current. The results are shown in Fig.~\ref{D_Gamma} for different edge lengths $l$ and different thicknesses of the sample. It can be clearly seen that $D_{\Gamma}$ is smaller than $D_0$. Using the original Thiele equation for the description of the vortex motion $D_{\Gamma}$ and $D_0$ are assumed to be equal.

Figures~\ref{fit1} and \ref{fit2} show an example for the fit of a numerically calculated vortex-core trajectory using both theories. The theory based on the modified Thiele equation shows better accordance than the theory based on the original Thiele equation. It can be seen that the Thiele equation has to be modified for a sufficient description of the dynamics of current-driven magnetic vortices in the presence of a non-adiabatic spin torque. This modification takes the deformation of the outer part of the vortex into account.

\end{document}